\documentclass[11pt]{article}

\usepackage{amssymb}
\usepackage{amsmath}

\usepackage{epsfig}
\usepackage{graphicx}
\usepackage{cite}
\usepackage{verbatim}
\usepackage{subfigure}
\usepackage{multicol}
\usepackage{picinpar}
\usepackage{pstricks}
\usepackage{pst-node}

\newtheorem{theorem}{\bf Theorem}
\newtheorem{observation}{\bf Observation}

\def\QED{\mbox{\rule[0pt]{1.5ex}{1.5ex}}}
\def\proof{\noindent\hspace{2em}{\it Proof: }}

\begin{document}
\setcounter{page}{1}
\title{{Capacity with Causal and Non-Causal Side Information\\ - A Unified View\footnote{This work was presented in part at the IEEE Communication Theory Workshop, June 12-15, 2005 and at the Forty-third Annual Allerton Conference on Communication, Control, and Computing, Sept. 28-30, 2005.}}}
\author{\normalsize Syed Ali Jafar\\
\small Center for Pervasive Communications and Computing\\
\small Department of Electrical Engineering and Computer Science\\ 
\small University of California Irvine, CA 92697 USA\\
{\small E-mail~:~ syed@ece.uci.edu}}
\date{}
\maketitle
\thispagestyle{empty}
\begin{abstract}We identify the common underlying form of the capacity expression that is applicable to both cases where causal or non-causal side information is made available to the transmitter. Using this common form we find that for the single user channel, the multiple access channel, the degraded broadcast channel, and the degraded relay channel, the sum capacity with causal and non-causal side information are identical when all the transmitter side information is also made available to all the receivers. A genie-aided outerbound is developed that states that when a genie provides $n$ bits of side information to a receiver the resulting capacity improvement can not be more than $n$ bits. Combining these two results we are able to bound the relative capacity advantage of non-causal side information over causal side information for both single user as well as various multiple user communication scenarios. Applications of these capacity bounds are demonstrated through examples of random access channels. Interestingly, the capacity results indicate that the excessive MAC layer overheads common in present wireless systems may be avoided through coding across multiple access blocks. It is also shown that even one bit of side information at the transmitter can result in unbounded capacity improvement. 
\end{abstract}

\newpage
\section{Introduction}
Characterizing the capacity of reliable communication between a transmitter and a receiver in the presence of side information is a problem as old as information theory itself. Starting with Shannon's characterization of the capacity with causal side information \cite{Shannon1958} and Kusnetsov and Tsybakov's seminal work \cite{Kusnetsov_Tsybakov} on coding for computer memories with defective cells that lead to a general characterization of capacity with non-causal side information by Gel'fand and Pinsker \cite{Gelfand_Pinsker}, the theory of communication with side information has evolved as a dichotomy between the two cases of non-causal side information and causal side information. Capacity results with non-causal side information were generalized to the case of rate limited side information by Heegard and El Gamal \cite{Heegard_Gamal} and more recently by Rosenzweig et. al. \cite{Rosenzweig_Steinberg_Shamai} for single user communication and by Cemal and Steinberg \cite{Cemal_Steinberg} for the multiple access channel. Study of non-causal side information has lead to interesting duality relationships such as that between source coding and channel coding with side information \cite{Cover_Chiang} as well as the duality between the Gaussian multiple input multiple output (MIMO) broadcast and multiple access channels \cite{Jindal_Vishwanath_Goldsmith, Dimacs2, mimo_bc_journal}. In addition to data storage \cite{Heegard_Gamal, Kusnetsov_Tsybakov} and data-hiding/watermarking/steganography \cite{Steinberg_Merhav} these results have found applications on Gaussian channels with additive Gaussian interference \cite{Costa,Gelfand_Pinsker,Kim_Sutivong_Sigurjonsson,Steinberg_Shamai_BC} recently leading to the determination of first the sum capacity \cite{Yu_Cioffi,Viswanath_Tse,Dimacs2,mimo_bc_journal} and then the entire capacity region for the MIMO broadcast channel \cite{Dimacs, DimacsViswanath, Weingarten_Steinberg_Shamai}. 

The case of causal side information has been studied separately by researchers \cite{Shannon1958,Salehi,Caire_Shamai_CSI,Das_Narayan,Sigurjonsson_Kim}. The original capacity result due to Shannon \cite{Shannon1958} requires coding over an extended alphabet of mappings from the channel state to the input alphabet. Caire and Shamai \cite{Caire_Shamai_CSI} showed that when the side information at the transmitter is a deterministic function of the side information available to the receiver, capacity achieving codes can be constructed directly on the input alphabet. The capacity of the time varying multiple access channel with causal side information was explored by Das and Narayan \cite{Das_Narayan} and more recently by Kim and Sigurjonsson \cite{Sigurjonsson_Kim}.

With the availability of all the above mentioned capacity results on causal and non-causal side information, there is a need to develop a unified view of communication with side information. A unified view would allow us to relate, combine and extend the existing results to new applications. In this paper, we approach this objective through the following questions:

\begin{enumerate}
\item {\it Common Framework: } What is the fundamental connection between the capacity characterizations with causal and non-causal side information at the transmitter?
\item {\it Value of a bit of side information:} What is the maximum possible capacity improvement with  $n$ bits of any kind (causal, non-causal, memoryless or correlated) of side information provided by a genie to the transmitter/receiver?
\item {\it Non-causal versus Causal:} What is the relative capacity advantage of non-causal side information over causal side information?
\item {\it Impact of Correlation:} How does correlation, either temporal or between the transmitter and receiver side information, affect the capacity?
\item {\it Multiple users:} Finally, how do the causal and non-causal capacity characterizations extend to multiple user channels, e.g. the multiple access, broadcast and relay channels?
\end{enumerate}
In this work we focus on both causal and non-causal side information, the relationship between them, and their extensions to multiuser communications. Our interest is in general capacity expressions with finite states. Specialized expressions for AWGN or fading channels may be obtained as special cases from these general expressions, subject to input distribution optimizations. 
\section{Background and Channel Model}
\begin{figure}[h]
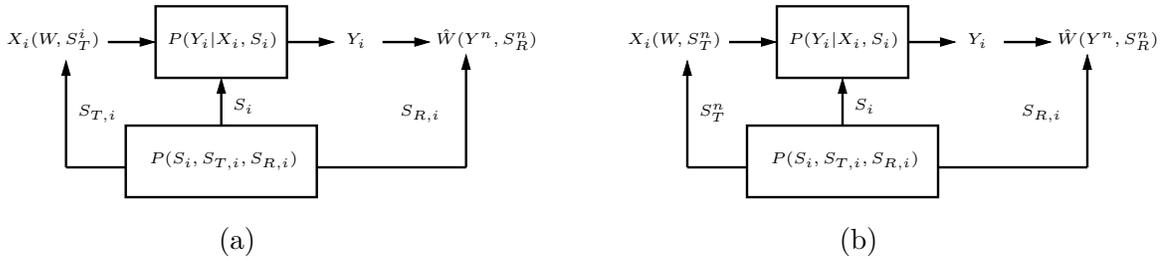

\[\begin{array}{ccc}
\input{channel_model_causal.pstex_t}&~~~~~~~~~~~&\input{channel_model_noncausal.pstex_t}\\
\mbox{(a)} &~& \mbox{(b)}\\
\end{array}
\]\caption{Channel Model with (a) Causal, and (b) Non-causal side information.}\label{fig:channel_model}
\end{figure}

The channel is a discrete memoryless channel (DMC) with message $W$, input $X_i$, output $Y_i$,  state $S_i$,  transmitter side information $S_{T,i}$ and receiver side information $S_{R,i}$ so that $P(S^n,S_T^n,S_R^n)=\Pi_{i=1}^n P(S_i,S_{T,i},S_{R,i})$ and  $P(Y^n|X^n,S^n)=\Pi_{i=1}^nP(Y_i|X_i,S_i)$. The only difference between the causal and non-causal side information cases is that with causal side information the input to the channel at time $i$ can only depend on the present and past states but not the future states, $X_i(W,S_T^i)$,  whereas with non-causal side information all inputs can depend on the entire state sequence $X_i(W,S_T^n)$. Notice that for the receiver it does not matter if the side information is made available causally or non-causally. This is because the receiver can wait till the end of transmission to decode the message. Figure \ref{fig:channel_model} illustrates the two scenarios. Probability of error, achievable rates and the capacity of this channel are defined in the standard sense \cite{Cover_Thomas, Caire_Shamai_CSI}.

For non-causal side information at the transmitter, the single user capacity is known to be \cite{Gelfand_Pinsker,Kusnetsov_Tsybakov,Heegard_Gamal}:
\begin{eqnarray}
C^{\mbox{\tiny non-causal}} = \max_{{\mathcal{P}}_{\mbox{\tiny non-causal}}} I(U;Y) - I(U;S_T)\label{CnonC}
\end{eqnarray}
where ${\mathcal{P}}_{\mbox{\tiny non-causal}}=\{P(U,X|S_T) =P(U|S_T)P(X|U,S_T)\}$. 
Comparing this to the case where no side information is available ($S_T=\phi$),
\begin{eqnarray}
C = \max_{P(U,X)} I(U;Y) = \max_{P(X)}I(X;Y), \label{CnoS}
\end{eqnarray}
note that the availability of side information at the transmitter is helpful in that the transmitter can match its input to the channel information by picking the input alphabet $U,X$ conditioned on $S_T$, as opposed to (\ref{CnoS}) where the input can not be matched to the channel state. However, the benefit of matching the input to the channel state comes with the cost of the subtractive term in (\ref{CnonC}), i.e., $I(U;S_T)$ which can be interpreted as the overhead required to communicate to the receiver, the adaptation to the channel state at the transmitter. 
For the case where the side information is available at the transmitter only causally, the capacity expression has been found by Shannon as \cite{Shannon1958}
\begin{equation}
C^{\mbox{\tiny causal}}  = \max_{P(t)}I(T;Y)\label{CcS}
\end{equation}
where $T$ is an extended alphabet of mappings from the channel state to the input alphabet. 

The capacity expressions (\ref{CnonC}), (\ref{CnoS}), (\ref{CcS}) explicitly account for side information at the transmitter. Side information at the receiver, $S_R$, is easily incorporated into the same expressions by replacing $Y$ with $(Y, S_R)$ in the corresponding expressions.

We start by finding a common form for the causal and non-causal cases.
\section{Relating Capacity with Causal and Non-causal Side Information}
Comparing the non-causal case (\ref{CnonC}) with the causal case (\ref{CcS}) the two capacity expressions are in different forms so that their relationship is not obvious. We start by making the relationship clearly apparent by representing the causal case in a different form, comparable to $(\ref{CnonC})$. The following result has also been observed and derived independently in parallel work by \cite{Sigurjonsson_Kim} and \cite{Cemal_Steinberg}.\\
\begin{eqnarray*}
C^{\mbox{\tiny non-causal}} &=& \max_{{\mathcal{P}}_{\mbox{\tiny non-causal}}}I(U;Y,S_R) - I(U;S_T),\\ 
C^{\mbox{\tiny causal}} &=& \max_{{\mathcal{P}}_{\mbox{\tiny causal}}} I(U;Y,S_R) - I(U;S_T),  \\
\end{eqnarray*}
with
\begin{eqnarray*}
{\mathcal{P}}_{\mbox{\tiny non-causal}}&=&\{P(U,X|S_T) = P(U|S_T)P(X|U,S_T)\}\\
{\mathcal{P}}_{\mbox{\tiny causal}}&=&\{P(U,X|S_T) = P(U)P(X|U,S_T)\}
\end{eqnarray*}
In the non-causal case, the choice of $U$ can be made conditional on the channel state $S_T$. In the causal case $U$ is picked independent of $S_T$.  This makes the subtractive term equal to zero for the causal case. In both cases, it suffices for the optimal input symbol $X$ to be just a function of $U,S_T$, i.e. $P(X|U,S_T)$ is either 0 or 1. 

{\it Sketch of Proof:} [Converse] Achievability of the capacity expression with causal side information is straightforward. Interestingly, the converse allows a very simple proof, as follows. Starting with Fano's inequality,
\begin{eqnarray}
nR&\leq& I(W;Y^n,S_R^n)+n\epsilon_n\\
&\leq& \sum_{i=1}^n I(W, S_T^{i-1};Y_i,S_{R,i}|Y^{i-1}, S_R^{i-1})\\
&=& \sum_{i=1}^n H(Y_i,S_{R,i} | Y^{i-1},S_R^{i-1})-H(Y_i,S_{R,i}| Y^{i-1},S_R^{i-1}, W, S_T^{i-1})\\
&\leq& \sum_{i=1}^n H(Y_i,S_{R,i})-H(Y_i,S_{R,i}| S_T^{i-1} ,W)\label{eq:presentpast}\\
&=& \sum_{i=1}^n I(U_i;Y_i,S_{R,i})
\end{eqnarray}
where $(\ref{eq:presentpast})$ follows because the current output is independent of the past outputs, conditioned on all the past inputs. $U_i=(W,S_T^{i-1})$ is the auxiliary random variable, independent of current channel state $S_{T,i}$.\hfill\QED

The capacity expression (\ref{CnonC}) has been shown \cite{Cover_Chiang,Chiang_Cover} to be the common form of single user capacity for all four cases of non-causal side information as well as the corresponding cases for rate-distortion.  In other words, for the capacity problem, whether the non-causal side information is available at the transmitter, the receiver, both, or neither, the capacity expression has the common form $I(U;Y, S_R)-I(U;S_T)$. The only difference is in the constraints on the distribution of the auxiliary random variable $U$, the input alphabet $X$ and the state variable $S$. Thus, combining the results of \cite{Cover_Chiang, Chiang_Cover} with the common expression obtained above we find that the expression $I(U;Y,S_R)-I(U;S_T)$ is indeed the common expression for not only all cases of non-causal side information but also for causal side information as well.
\[\begin{array}{lll}
C^{\mbox{\tiny non-causal}}(S_T, S_R) = \max_{P(U,X|S_T)} & I(U; Y, S_R)-I(U;S_T) &\\
C^{\mbox{\tiny non-causal}}(\phi, S_R) = \max_{U=\phi, P(X)} & I(U; Y, S_R)-I(U;S_T) &= \max_{P(X)} I(X;Y,S_R) \\
C^{\mbox{\tiny non-causal}}(S_T, \phi) = \max_{P(U,X|S_T)} & I(U; Y, S_R)-I(U;S_T) &= \max_{P(U,X|S_T)} I(U;Y) - I(U;S_T)\\
C^{\mbox{\tiny non-causal}}(\phi, \phi) = \max_{U=\phi,P(X)} & I(U; Y, S_R)-I(U;S_T) &=\max_{P(X)} I(X;Y)\\
\end{array}
\]
\[\begin{array}{lll}
C^{\mbox{\tiny causal}}(S_T, S_R) = \max_{P(U)P(X|U, S_T)} & I(U; Y, S_R)-I(U;S_T) &= \max_{P(U)P(X|U, S_T)}I(U;Y,S_R)\\
C^{\mbox{\tiny causal}}(\phi, S_R) = \max_{U=\phi, P(X)} & I(U; Y, S_R)-I(U;S_T) &= \max_{P(X)} I(X;Y,S_R) \\
C^{\mbox{\tiny causal}}(S_T, \phi) = \max_{P(U)P(X|U,S_T)} & I(U; Y, S_R)-I(U;S_T) &= \max_{P(U)P(X|U,S_T)} I(U;Y) \\
C^{\mbox{\tiny causal}}(\phi, \phi) = \max_{U=\phi, P(X)} & I(U; Y, S_R)-I(U;S_T) &= \max_{P(X)} I(X;Y)\\
\end{array}
\]

Investigating the relationship between causal and non-causal information further, we prove that the two capacities are the same if the transmitter side information is also available to the receiver.

\begin{theorem}\label{theorem:cnc} \emph{(Relationship between causal and non-causal side information capacity)}
{If the side-information at the transmitter is a deterministic function of the side-information at the receiver, i.e., if $S_T=f(S_R)$,  then capacity with causal side information is equal to the capacity with non-causal side information.} 
\end{theorem}
Capacity achieving codes, in both cases, can be constructed directly on the input alphabet and the auxiliary random variable $U$ is not required. \\

\proof
\begin{eqnarray*}
C^{\mbox{\tiny non-causal}} &=& \max_{P(U,X|S_T)=P(U|S_T)P(X|U,S_T)} I(U;Y, S_R) - I(U;S_T)\\
&=& \max_{P(U,X|S_T)=P(U|S_T)P(X|U,S_T)} I(U;S_R) + I(U;Y|S_R)-I(U;S_T)\\
&=& \max_{P(U,X|S_T)=P(U|S_T)P(X|U,S_T)} I(U;S_R,S_T)+I(U;Y|S_R)-I(U;S_T)\\
&=& \max_{P(U,X|S_T)=P(U|S_T)P(X|U,S_T)} I(U;S_R|S_T)+I(U;Y|S_R)\\
&=& \max_{P(U,X|S_T)=P(U|S_T)P(X|U,S_T)} I(U;Y|S_R)\\
&=& \max_{P(X|S_T)} I(X;Y|S_R)\\
&=& C^{\mbox{\tiny causal}}
\end{eqnarray*}
where the last equality follows from the results of \cite{Caire_Shamai_CSI} for causal side information. It is shown in \cite{Caire_Shamai_CSI} that with causal side information at the transmitter, if $S_T=f(S_R)$, i.e., if the side-information at the transmitter is a deterministic function of the side-information at the receiver then the auxiliary random variable $U$ is not needed and coding can be performed directly on the input alphabet $X$.

Thus, if the side information at the transmitter is a deterministic function of the side information at the receiver, then capacity with non-causal side information is the same as the capacity with causal side-information. \hfill\QED

Next we investigate the value of side information through capacity bounds.

\def\geniebceq{(Y,Z)=\left\{\begin{array}{ll}{(aX+N, 0)},& {G=0}\\ {(0, aX+N)}, &{G=1}\end{array}\right.}
\def\channeleq{p(H~|~U,V)}

\section{Genie bits and the Value of Side Information}
In this section we answer the question: what is the maximum capacity gain that can result from a fixed number of bits of side information? For example, suppose a genie provides a fixed number of bits of side information $G$ to the transmitter or the receiver per channel use. There are no constraints on the genie provided side information, i.e., it could be causal, non-causal, temporally correlated, or i.i.d. A fundamental question is whether the capacity increase, $C_G-C$, due to a fixed number of genie bits is bounded and if so, then what is the maximum capacity benefit. The following results show that transmitter and receiver side information are fundamentally different in their potential capacity advantages. 
\subsection{Receiver Side Information}
\begin{theorem} \label{theorem:genie}The maximum possible capacity improvement due to the availability of receiver side information is bounded by the amount of the side information itself.
\begin{eqnarray*}
C_G-C\leq H(\mathcal{G})=\lim_{N\rightarrow\infty}\frac{1}{N}H(G_1,G_2,\cdots,G_N).
\end{eqnarray*}
\end{theorem}


\proof
\begin{eqnarray*}
C_G&=&\sup_{p()}\lim_{N\rightarrow\infty}\frac{1}{N}I(W;Y^N,G^N)\\
&=&\sup_{p()}\lim_{N\rightarrow\infty}\frac{1}{N}\left(I(W;Y^N)+I(W;G^N|Y^N)\right)\\
&=& C + \Delta C
\end{eqnarray*}
where the $\sup$ is over the multi-letter distributions of allowed input strategies. $C_G$ is the  capacity with the side information provided by the genie, $C$ is the capacity without the side information and $\Delta C$, the capacity improvement, is bounded by the entropy rate $H({\mathcal{G}})$. In other words, if the genie provides one bit of side information per channel use, the capacity benefit $\Delta C=C_G-C$ can not be more than 1 bit, regardless of the kind of side information. \hfill\QED

\subsection{Transmitter Side Information}
\begin{theorem}\label{theorem:unbounded}
The maximum possible capacity improvement from a fixed number of genie bits (per channel use) of side information is unbounded, even if the side information is causal.
\end{theorem}
\proof The proof is in the form of the following example. Consider the channel with input alphabet $X\in\mathcal{X}=\{0,1,2,\cdots, 2N-1\}$, output alphabet $Y\in\mathcal{Y}=\mathcal{X}\cup\{\phi\}$ where $\phi$ is the erasure symbol, and i.i.d. state sequence drawn from the alphabet $S\in\mathcal{S}=\{0,1\}$. The input output relationship is given by,
\begin{eqnarray*}
Y&=&\left\{\begin{array}{ll} 
X& \mbox{when }  X+S \mbox{ is even,}\\
\phi&\mbox{when }  X+S \mbox{ is odd.}
\end{array}
\right.
\end{eqnarray*}
In simple words, when the channel state is $S=0$, the channel conveys even inputs noise-free and erases odd inputs. Similarly, when the channel state is $S=1$, the channel conveys odd inputs noise-free and erases even inputs. The channel state $S$ is unknown to the receiver.

\begin{figure}[h]
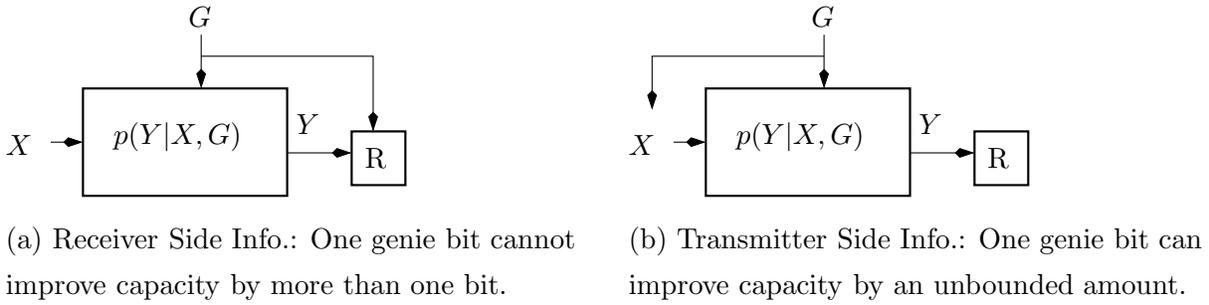

\[\begin{array}{lcl}
\input{genieRx.pstex_t}&&\input{genieTx.pstex_t}\\
\mbox{(a) Receiver Side Info.: One genie bit cannot} && \mbox{(b) Transmitter Side Info.: One genie bit can }\\
\mbox{improve capacity by more than one bit.} && \mbox{improve capacity by an unbounded amount.}\\
\end{array}
\]
\caption{Value of Side Information at the Transmitter and Receiver}\label{fig:genie}
\end{figure}

Suppose the genie provides one bit of side information in the form of $G=S$ to the transmitter every channel use. With perfect knowledge of $S$ at the transmitter, the capacity of this channel is clearly $C_G=\log(N+1)$ where $N+1$ is the number of distinct outputs that can be affected by the transmitter. On the other hand, with no state information at the transmitter the optimal input distribution is uniform on $\mathcal{X}$, the corresponding output distribution is 
\begin{eqnarray*}
Y&=&\left\{\begin{array}{ll} 
\phi & \mbox{with probability } \frac{1}{2}\\
i& \mbox{with probability } \frac{1}{4N}, ~~i\in\mathcal{X}
\end{array}
\right.
\end{eqnarray*}
 and the capacity is
\begin{eqnarray*}
C=I(X;Y)&=& H(Y)-H(Y|X) = \frac{1}{2}\log(N)+\frac{1}{2}.
\end{eqnarray*}
The capacity benefit of one genie bit therefore is 
\begin{eqnarray*}
C_G-C=\log(N+1)-\frac{1}{2}\log(N)-\frac{1}{2}\geq\frac{1}{2}\log(N)-\frac{1}{2}
\end{eqnarray*}
which is unbounded, i.e. goes to infinity as $N$ goes to infinity. Notice that $C_G$ is obtained with only causal side information at the transmitter. Thus, the example shows that the capacity benefit of one bit of side information at the transmitter can be unbounded even when the side information is causal.\hfill\QED\\
\noindent{The} contrasting potential capacity benefits of side information are summarized in Fig. \ref{fig:genie}.


While we have shown that, in theory, unbounded capacity improvement can result from a finite amount of causal side information at the transmitter, it is not clear how such side information and the associated capacity benefits can be obtained in practice. For practical systems, side information at the transmitter is often obtained through a feedback channel from the receiver. However, it is well known that for a DMC, even if all the past channel outputs obtained at the receiver are made available to the transmitter through a noise and delay free feedback channel, the capacity is unaffected. Thus, the unbounded potential capacity benefits of causal side information are in sharp contrast with the zero capacity benefit of causal feedback for a DMC. The key to reconciling these results lies in the definition of causal in both cases. While causal side information allows the transmitter knowledge of the current channel state, causal feedback allows information only about the past channel outputs. For a memoryless channel past outputs do not provide any information about the current channel state. In the example considered above, causal feedback of the channel outputs will provide the transmitter precise knowledge of all the past channel states, but no information about the present channel state. Thus, the timing of the side information at the transmitter can make the difference between unbounded capacity improvements and no improvement at all.

\section{Advantage of Non-Causal Side Information over Causal Side Information}
The preceding section shows that the present state information at the transmitter can be invaluable compared to the information of all the past states. The natural question then is, what is the advantage of knowing the \emph{future} channel states over knowing just the present and past states? In other words, what is the capacity advantage of non-causal side information over causal side information? It turns out, the answer to this question is already implicit in the results of Theorems \ref{theorem:cnc} and \ref{theorem:genie}.
\begin{theorem}\label{theorem:nc-c}
The capacity benefit of non-causal side information over causal side information is bounded as follows:
\begin{eqnarray*}
C^{\mbox{\tiny non-causal}}(S_T,S_R)-C^{\mbox{\tiny causal}}(S_T,S_R)&\leq& H(S_T|S_R)
\end{eqnarray*}
\end{theorem}

\proof
From the preceding sections we have the following two results.
\begin{itemize}
\item If the transmitter side information is also available to the receiver, capacity with causal and non-causal side information is identical. (Theorem \ref{theorem:cnc})
\item If a genie provides a bit of side information to the receiver it can not improve capacity by more than a bit. (Theorem \ref{theorem:genie})
\end{itemize}
Combining these two results, suppose the transmitter side information is made available to the receiver by a genie. This requires $H(S_T|S_R)$ genie bits and therefore cannot improve capacity by more than $H(S_T|S_R)$ bits. Using the results stated above, we have:
\begin{eqnarray*}
C^{\mbox{\tiny non-causal}}(S_T,(S_T,S_R))-H(S_T|S_R)&\leq& C^{\mbox{\tiny non-causal}}(S_T,S_R)\leq C^{\mbox{\tiny non-causal}}(S_T,(S_T,S_R))\\
C^{\mbox{\tiny causal}}(S_T,(S_T,S_R))-H(S_T|S_R)&\leq& C^{\mbox{\tiny causal}}(S_T,S_R)\leq C^{\mbox{\tiny causal}}(S_T,(S_T,S_R))\\
C^{\mbox{\tiny non-causal}}(S_T,(S_T,S_R))&=&C^{\mbox{\tiny causal}}(S_T,(S_T,S_R))\\
C^{\mbox{\tiny non-causal}}(S_T,S_R)-C^{\mbox{\tiny causal}}(S_T,S_R)&\leq& H(S_T|S_R)
\end{eqnarray*}
In simple words, the advantage of non-causal side information over causal side information is bounded by the number of genie bits required to make the transmitter side information available to the receiver as well.
\subsection{Example: Random Access}
Consider a single user DMC characterized by $P(Y|X)$ and with a capacity $C_0 = \max_{P(X)}I(X;Y)$ when the input is directly controlled by a transmitter. Now, suppose instead that the transmitter is only able to access the channel (control the input) in a random manner as:
\begin{equation*}
X= X_TS+X_r(1-S),
\end{equation*}
where $S\in\{0,1\}$ is a switch state that determines when the transmitter can access the channel with the symbol $X_T$, and $X_r$ is a randomly generated input. 

\begin{figure}[h]
\centerline{\input{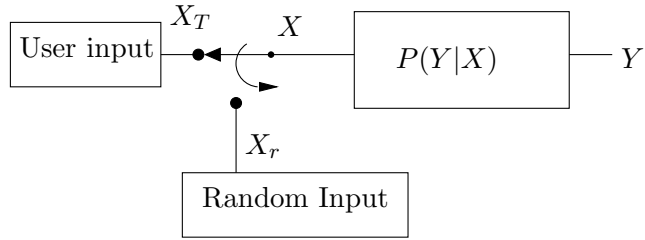}}
\caption{Random Access Channel}\label{fig:randomaccess}
\end{figure}

Suppose the state $S$ is known to the transmitter and \emph{not known} to the receiver. Such a channel is  relevant to cognitive communication scenarios \cite{Syed_cognitive} and is also similar to the "memory with stuck-at defects" problem considered in \cite{Heegard_Gamal}. Clearly, if the switch state is provided to the receiver by a genie the resulting capacity is $C(S,S)=\mbox{Prob}(S=1)C_0=\overline{S}C_0$. Since the extra information provided by the genie is only one bit we have
\begin{eqnarray*}
\overline{S}C_0&\geq& C^{\mbox{\tiny non-causal}}(S,\phi)\geq \overline{S}C_0-1
\end{eqnarray*}
Interestingly, the capacity with causal side information in this case is the same as the capacity with no side information. This is easily seen by rewriting $X_T=f(U,S)$ as follows:
\begin{equation*}
X_T=\left\{\begin{array}{l} f_0(U), S=0,\\ f_1(U), S=1
\end{array}\right. \equiv f_1(U), S=0,1.
\end{equation*}
In other words, the choice of input symbol does not matter when the switch is open $(S=0)$. Thus, we have
\begin{equation*}
C^{\mbox{\tiny causal}}(S,\phi)=C(\phi,\phi)\geq C(S,S)-1.
\end{equation*}
The effect of memory in side information is also revealed by this example. Suppose the switch changes state in a block static model, i.e. it retains its state for $N$ symbols and then changes to an i.i.d. realization. In this case, the genie only needs to provide one bit to the receiver every $N$ channel uses and the bounds are tighter.
\begin{equation*}
\overline{S}C_0=C(S,S)\geq C^{\mbox{\tiny non-causal}}(S,\phi)\geq C^{\mbox{\tiny causal}}(S,\phi)=C(\phi,\phi)\geq\overline{S}C_0-\frac{1}{N}
\end{equation*}
\section{Multiple Access Channel with Independent Side Information}
Achievable regions with causal side information are straightforward to obtain because the codewords can always be constructed on mappings from the side information to the channel input alphabet. In the multiple access channel, the two transmitters have (possibly correlated) side information $S_{T1}, S_{T2}$ respectively, and the common receiver has side information $S_{R}$. The characterization of the achievable region with side information is the same as without side information, with the codes defined on auxiliary random variables that are independent of the side information, and the actual channel input symbols chosen as a function of the auxiliary random variable and the instantaneous side information. Thus the following achievable region is obtained.
\begin{eqnarray*}
R_1&\leq &I(U_1;Y,S_R|U_2) = I(U_1;Y|U_2,S_R)\\
R_2&\leq &I(U_2;Y,S_R|U_1) = I(U_2;Y|U_1,S_R)\\
R_1+R_2&\leq &I(U_1,U_2;Y,S_R) = I(U_1,U_2;Y|S_R)
\end{eqnarray*}
where $U_1, U_2$ are mutually independent as well as independent of the side information and the channel inputs are given by $X_1=f_1(U_1,S_{T1}), X_2=f_2(U_2,S_{T2}).$

While a full converse is not known, upperbounds for the MAC with causal side information are obtained in \cite{Sigurjonsson_Kim} in terms of the capacity achieved with transmitter cooperation. In this work, we focus on the multiple access channel with independent side information at the two transmitters. We prove that the sum capacity of the multiple access channel with independent side information is given by the corresponding constraint in the achievable region provided above.
\begin{theorem}
The sum capacity of the discrete memoryless multiple access channel with causal side information $S_{T1}, S_{T2}$ ($S_{T1},S_{T2}$ independent) and $S_R$ available to transmitter 1, transmitter 2 and the receiver, respectively, is given by
\begin{equation}
R_1+R_2= \max I(U_1,U_2;Y,S_R)
\end{equation}
with $P(U_1,U_2,X_1,X_2,S_{T1},S_{T2})=P(U_1)P(U_2)P(X_1|U_1,S_{T1})P(X_2|U_2,S_{T2})P(S_{T1},S_{T2})$.
\end{theorem}
\proof The converse is proved as follows:
\begin{eqnarray*}
n(R_1+R_2)&\leq & I(W_1,W_2;Y^n,S_R^n)+n\epsilon\\
&\leq & \sum_{i=1}^n I(W_1,W_2,S_{T1}^{i-1},S_{T2}^{i-1};Y_i,S_{R,i}|Y^{i-1},S_R^{i-1})+n\epsilon\\
&\leq & \sum_{i=1}^n H(Y_i,S_{R,i})-H(Y_i,S_{R,i}|W_1,S_{T1}^{i-1},W_2,S_{T2}^{i-1},Y^{i-1},S_R^{i-1})+n\epsilon\\
&=&\sum_{i=1}^n I(U_{1,i},U_{2,i};Y_i,S_{R,i})+n\epsilon
\end{eqnarray*}
where $U_{1,i}=W_1,S_{T1}^{i-1}$ and $U_{2,i}=W_2,S_{T2}^{i-1}$ are independent of $S_{T1,i}, S_{T2,i}$. Independence of messages and transmitter side information implies independence of the auxiliary random variables $U_1$, $U_2$ as well.
\hfill\QED

For non-causal side information, an achievable region is readily obtained when the side information at the two transmitters is independent.

\begin{eqnarray*}
C\supseteq\left\{(R_1,R_2)\right.&:& R_1\leq I(U_1;Y,S_R|U_2)-I(U_1;S_{T1})\\
&& R_2\leq I(U_2;Y,S_R|U_1) - I(U_2;S_{T2})\\
&&R_1+R_2\leq\left. I(U_1, U_2;Y,S_{R})-I(U_1;S_{T1})-I(U_2;S_{T2})\right\}
\end{eqnarray*}
for all  $P(U_1,X_1,U_2,X_2|S_{T1},S_{T2})=P(U_1,X_1|S_{T1})P(U_2,X_2|S_{T2})$.
However, to the best of the author's knowledge a converse has not been shown  for independent side information\footnote{The problem with extending the single user approach appears to be that including $Y^{i-1}$ into the auxiliary random variables makes them correlated.}. Correlation of the side information makes even the achievable region non-trivial, as the possibility of Slepian Wolf coding of correlated side information at the two transmitters can be exploited.

For our purpose however, we show that with independent side information at the transmitters, if all the transmitter side information is also made available to the receiver, i.e. $(S_{T1},S_{T2})=f(S_R)$, then the MAC capacity region with causal side information is identical to the capacity region with non-causal side information.
\begin{theorem}\label{theorem:mac_cnc}
For the discrete memoryless multiple access channel with side information $S_{T1}, S_{T2}$ (independent) and $S_R$ available to transmitter 1, transmitter 2 and the receiver, respectively, if $(S_{T1}, S_{T2})=f(S_{R})$ then the capacity region for both causal or non-causal side information is given by the convex hull of all rates pairs $(R_1, R_2)$ satisfying the following inequalities:
\begin{eqnarray*}
R_1&\leq& I(X_1;Y|S_R,X_2)\\
 R_2&\leq& I(X_2;Y|S_R,X_1)\\
R_1+R_2&\leq&\left. I(X_1, X_2;Y|S_{R})\right\}
\end{eqnarray*}
for all $P(X_1|S_{T1}), P(X_2|S_{T2})$. 
\end{theorem}
\proof [Achievability]
Achievability is easily established as follows. Starting with the achievable region for the causal side information case, we have:
\begin{eqnarray}
R_1&\leq& I(U_1;Y|U_2, S_R)\\
&=& H(Y|U_2,S_R)-H(Y|U_2,U_1,S_R)\\
&=& H(Y|U_2,S_R,X_2)-H(Y|U_2,U_1,S_R,X_2,X_1) \label{eq:X}\\
&=& H(Y|X_2,S_R)-H(Y|X_1,X_2,S_R)\\
&=& I(X_1;Y|S_R,X_2).
\end{eqnarray}
Equation (\ref{eq:X}) follows from the fact that $X_1$ (resp. $X_2$) is a function of $U_1, S_{T1}$ (resp. $U_2, S_{T2}$) and $S_{T1}$ (resp. $S_{T2}$) is a function of $S_R$. Thus, $X_1$ (resp. $X_2$) is a function of $U_1, S_R$ (resp. $U_2, S_R$). The corresponding inequalities for $R_2$ and the sum rate $R_1+R_2$ are similarly obtained. Clearly, what is achievable with causal side information is also achievable with non-causal side information.\hfill\QED

\proof [Converse]
For the converse, we start with the individual rate constraints. For both causal and non-causal side information we have:
\begin{eqnarray}
nR_1&\leq& I(W_1;Y^n,S_R^n|W_2)+n\epsilon\\
&=&\sum_{i=1}^n I(W_1;Y_i|W_2,Y^{i-1},S_R^n)+n\epsilon\\
&=&\sum_{i=1}^n H(Y_i|S_{R}^n,W_2,Y^{i-1},X_{2,i})-H(Y_i|W_2,W_1,S_R^n,Y^{i-1},X_{1,i},X_{2,i}) \label{eq:XX}\\
&\leq&\sum_{i=1}^n H(Y_i|S_{R,i},X_{2,i})-H(Y_i|W_2,W_1,S_R^n,Y^{i-1},X_{1,i},X_{2,i})\label{eq:XXX}\\
&=& \sum_{i=1}^n H(Y_i|S_{R,i},X_{2,i})-H(Y_i|X_{1,i},X_{2,i},S_{R,i})+n\epsilon\\
&=&\sum_{i=1}^nI(X_{1,i};Y_i|X_{2,i},S_{R,i})+n\epsilon
\end{eqnarray}
Independence of side information ensures that $X_{1,i}\rightarrow S_{1,i}\rightarrow S_{2,i}\rightarrow X_{2,i}$ form a Markov chain as required.

Equation (\ref{eq:XX}) follows from the fact that $X_{2,i}$ is a function of $W_2$ and $S_R^{n}$ (resp. $S_R^i$) for non-causal (resp. causal) side information. (\ref{eq:XXX}) follows because conditioning reduces entropy and because $Y_i$ is independent of $Y^{i-1}, W_1, W_2, S_R^{i-1}, S_{R,i+1}^n$ given $S_{R,i}, X_{1,i}, X_{2,i}$. Note that the inequality in (\ref{eq:XXX}) is necessary because $Y_i$ is not independent of $Y^{i-1},  W_2, S_R^{i-1}, S_{R,i+1}^n$ given $S_{R,i}, X_{2,i}$. Intuitively, $Y^{i-1}$ contains some information about $W_1$ and thus $X_{1,i}$ which affects $Y_i$. The converse for $R_2$ follows similarly. Finally, for the sum rate we have:
\begin{eqnarray*}
n(R_1+R_2)&\leq &I(W_1,W_2;Y^n,S_R^n)+n\epsilon\\
&=& I(W_1,W_2;Y^n|S_R^n)+n\epsilon\\
&=& \sum_{i=1}^n H(Y_i|S_R^n)-H(Y_i| W_1,W_2,Y^{i-1},S_R^n)+n\epsilon\\
&\leq& \sum_{i=1}^n H(Y_i|S_{R,i})-H(Y_i|X_{1,i},X_{2,i},S_{R,i})+n\epsilon\\
&\leq& \sum_{i=1}^n I(X_{1,i},X_{2,i};Y_i|S_{R,i})+n\epsilon
\end{eqnarray*}
\hfill\QED
\subsection{Genie Bits and Value of Side Information for a Multiple Access Channel}
Theorem \ref{theorem:mac_cnc} extends the single user result of Theorem \ref{theorem:cnc} to the multiple access channel. By analogy to the single user case, a relationship between causal and non-causal side information capacity can be obtained by extending the results of Theorem \ref{theorem:genie} to the multiple access channel. Recall that in Theorems \ref{theorem:genie} and \ref{theorem:unbounded} we  characterized the potential capacity benefits of side information at the transmitter and receiver for single user communications. Clearly, Theorem \ref{theorem:unbounded} extends directly to the multiple access channel, because if unbounded capacity gains are possible with transmitter side information in a single user channel, then the same must be true of the multiple access channel which includes the single user capacity as a special point in its capacity region. The extension of Theorem \ref{theorem:genie} to the multiple access channel is only slightly less straightforward as the derivation for the single user channel can be easily modified to the multiple access case as follows.
\begin{theorem} \label{theorem:macgenie} For the multiple access channel, the maximum possible sum capacity improvement $C^\Sigma_G-C^\Sigma$ due to the availability of receiver side information is bounded by the amount of the side information itself.
\begin{eqnarray*}
C^\Sigma_G-C^\Sigma\leq H(\mathcal{G})=\lim_{N\rightarrow\infty}\frac{1}{N}H(G_1,G_2,\cdots,G_N).
\end{eqnarray*}
\end{theorem}
\proof
\begin{eqnarray*}
C^\Sigma_G&=&\sup_{p(X_{1,1^N}(W_1),X_{2,1}^N(W_2))}\lim_{N\rightarrow\infty}\frac{1}{N}I(W_, W_2;Y^N,G^N)\\
&=&\sup_{p(X_{1,1^N}(W_1),X_{2,1}^N(W_2))}\lim_{N\rightarrow\infty}\frac{1}{N}\left(I(W_1, W_2;Y^N)+I(W_1,W_2;G^N|Y^N)\right)\\
&=& C^\Sigma + \Delta C^\Sigma
\end{eqnarray*}
where $C_G^\Sigma$ is the sum capacity with the side information provided by the genie, $C^\Sigma$ is the sum capacity without the side information and $\Delta C$, the capacity improvement, is bounded by the entropy rate $H({\mathcal{G}})$. Thus, the single user capacity result extends directly to the multiple access sum capacity. If the genie provides one bit of side information to the common receiver per channel use, the \emph{sum} capacity benefit $\Delta C=C_G-C$ can not be more than 1 bit, regardless of the kind of side information. \hfill\QED

\subsection{Advantage of Non-causal Side Information over Causal Side Information}
We compare the causal and non-causal capacity regions in terms of the sum rate point $C_\Sigma$. Similar to the single user case, we have shown that the sum capacities (and the entire capacity regions) are identical when $S_{T1}, S_{T2}$ (independent) are also available to the receiver. To make this information available to the receiver requires $H(S_{T1},S_{T2}|S_R)$ genie bits per symbol. And because we have shown for the multiple access channel that genie bits can not improve capacity by more than their own entropy, we have the following result:
\begin{theorem}\label{theorem:macnc-c}
For the multiple access channel with independent side information at the transmitters, the sum capacity benefit of non-causal side information over causal side information is bounded as follows:
\begin{eqnarray*}
C^{\mbox{\tiny non-causal}}_\Sigma(S_T,S_R)-C_\Sigma^{\mbox{\tiny causal}}(S_T,S_R)&\leq& H(S_{T1},S_{T2}|S_R)
\end{eqnarray*}
\end{theorem}
\proof
\begin{eqnarray*}
C_\Sigma(S_{T1},S_{T2},(S_R,S_{T1},S_{T2}))&\geq& C_\Sigma^{\mbox{\tiny non-causal}}(S_{T1},S_{T2},S_R)\\
&\geq& C_\Sigma^{\mbox{\tiny causal}}(S_{T1},S_{T2},S_R)\\
&\geq& C_\Sigma(S_{T1},S_{T2},(S_R,S_{T1},S_{T2}))-H(S_{T1},S_{T2}|S_R).
\end{eqnarray*}
\hfill\QED
\subsection{Example: Random Access Channel with Multiple Users}
The following example presents a scenario with correlated side information and shows how the results may be applicable in several such cases as well. Consider the random access channel described before, except now the channel input is controlled by two users as:
\begin{eqnarray*}
X=X_{1}S+X_{2}(1-S).
\end{eqnarray*}
\begin{figure}[h]
\centerline{\input{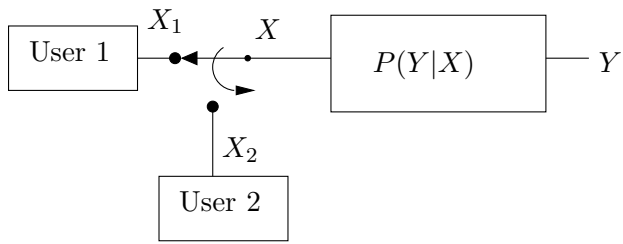}}
\caption{Random Access Channel with Two Users}\label{fig:macrandomaccess}
\end{figure}
Thus, a scheduler randomly allows user 1 or user 2 to access the channel in bursts of $N$ symbols. As before, the switch state $S$ is known to the transmitters but not to the receiver. If a genie provides $S$ to the receiver, the sum capacity is $C_0$. Using the same arguments as in the single user example, we have:
\begin{eqnarray*}
C_0&=&C_\Sigma(S,S,S)\\
&\geq& C_\Sigma^{\mbox{\tiny non-causal}}(S,S,\phi)\\
&\geq &C_\Sigma^{\mbox{\tiny causal}}(S,S,\phi)\\
&\geq &C_\Sigma(S,S,S)-\frac{1}{N}H(S)=C_0-1/N.
\end{eqnarray*}
The practical implications of this example are quite interesting. In practice, random access is handled by the medium access layer (MAC layer) through explicit handshakes between the transmitter and receiver in the form of RTS-CTS (request to send, clear to send) messages. For a rapidly varying random access channel such exchanges can constitute a major overhead. Practical wireless systems such as the 802.11 Wireless LANs use nearly half the resources just for MAC layer overheads. However, the capacity results presented above show that even in the extremely rapidly varying random access channels where a transmission may or may not be made each symbol period without the knowledge of the receiver, the capacity loss is limited to less than a bit per channel use. For random access that fluctuates at a less rapid scale, the loss is negligible. This has interesting implications in how the RTS-CTS overhead can be minimized in practical systems through coding across access attempts.
\section{The Broadcast Channel with Side Information}
In the general broadcast channel, the transmitter has side information $S_{T}$ and the two receivers have side information $S_{R1}$ and $S_{R2}$. The capacity region of the general broadcast channel is not known even without side information. The best known achievable region is due to Marton \cite{Marton} with auxiliary variables $U, V$ and $W$. When the transmitter has causal side information,  Marton's achievable region is directly extended to the convex hull of rate pairs $(R_1,R_2)$ satisfying:
\begin{eqnarray*}
R_1&\leq &I(W,U;Y_1, S_{R1}) = I(W,U;Y_1|S_{R1})\\
R_2&\leq &I(W,V;Y_2, S_{R2}) = I(W,V;Y_2|S_{R2})\\
R_1+R_2&\leq&\min\{I(W;Y_1, S_{R1})+I(W;Y_2,S_{R2})\}+I(U;Y_1,S_{R1}|W)+I(V;Y_2,S_{R2}|W)-I(U;V|W)\\
&=&\min\{I(W;Y_1| S_{R1})+I(W;Y_2|S_{R2})\}+I(U;Y_1|S_{R1},W)+I(V;Y_2|S_{R2},W)-I(U;V|W)
\end{eqnarray*}
where $P(U,V,W,X|S_{T})=P(U,V,W)P(X|S_T)$. Note that the only difference in the achievable region with causal side information versus Marton's innerbound without causal side information is that with causal side information the mapping from the auxiliary random variables $U,V,W$ to the channel input alphabet $X$ depends on the current value of the side information $S_T$, i.e. $X=f(U,V,W,S_T)$. In both cases $U,V,W$ are independent of the side information. As usual the achievability proof for causal side information is straightforward as the mapping $f(.)$ can be incorporated into the channel to obtain a case with no side information where Marton's innerbound applies.

Extension of Marton's innerbound to case where non-causal side information is available at the transmitter is less straightforward. A simple and elegant proof of achievability of a subset of Marton's innerbound with only the auxiliary random variables $U, V$ was provided by El Gamal and Van Der Meulen in \cite{Gamal_Meulen} \footnote{While their proof directly addresses only the random variables $U,V$ it can be extended to include $W$ and hence the entire Marton's innerbound region}. The achievable region with only $U,V$ is the set of rates $(R_1,R_2)$ satisfying
\begin{eqnarray*}
R_1 & \leq & I(U;Y_1)\\
R_2 & \leq & I(V;Y_2)\\
R_1+R_2 &\leq& I(U;Y_1)+I(V;Y_2)-I(U;V)
\end{eqnarray*}
for all $P(U,V,X)$. 

El Gamal and van der Meulen's simple achievability proof was extended by the author in \cite{Syed_CTW}  to incorporate non-causal side information as follows:
\begin{eqnarray*}
R_1 & \leq & I(U;Y_1,S_{R1})-I(U;S_T)\\
R_2 & \leq & I(V;Y_2,S_{R2})-I(V;S_{T})\\
R_1+R_2 &\leq& I(U;Y_1,S_{R1})+I(V;Y_2,S_{R2})-I(U;V) - I(U,V;S_T)
\end{eqnarray*}
for all $P(U,V,X|S_T)=P(U,V)P(X|U,V,S_T)$. While a full extension of Marton's innerbound with side information has been found recently, the proof of the abovementioned limited extension is interesting for its simplicity and also illustrates the binning concept central to the full extension. Since, the same concept can be applied to obtain achievable regions for many multiuser communication scenarios with non-causal side information \cite{Syed_CTW} we include a sketch of this proof here.

{\it [Sketch of Proof:]} Proceeding as in \cite{Gamal_Meulen}, $2^{n(I(U;Y_1,S_{R1})-\epsilon)}$ (resp. $2^{n(I(V;Y_2,S_{R2})-\epsilon)}$) i.i.d. $U$ sequences (resp. $V$ sequences) are generated independently according to $P(U)$ (resp. $P(V)$) and uniformly distributed over $2^{nR_1}$ (resp. $2^{nR_2}$) bins. This forms the codebook that is shared by all parties prior to beginning of communication. During the communication phase, the message index $W_1\in[1,2^{nR_1}]$  (resp. $W_2\in[1,2^{nR_2}]$) marks the appropriate $U$ (resp. $V$) sequence bin selected. Given the state sequence $S_T^n$, the transmitter finds a $U$ sequence and a $V$ sequence in the chosen bins so that $U^n, V^n, S^n$ are jointly typical. The probability that independently generated $U^n$ (resp. $V^n$) and $S^n$ are jointly typical is bounded by $2^{-n(I(U;S_T)-\delta(\epsilon))}$ (resp. $2^{-n(I(V;S_T)-\delta(\epsilon))}$). This is reflected in the individual rate constraints. Finally, the probability that independently generated $U^n, V^n$ and $S^n$ are jointly typical is bounded by $2^{-n(H(U)+H(V)+H(S)-H(U,V,S))-\delta(\epsilon)} = 2^{-n(I(U;V)+I(U,V;S))-\delta(\epsilon)}$ which results in the sum rate constraint.\hfill\QED

The complete extension of Marton's innerbound extension to non-causal side information was obtained in an independent and parallel work by Steinberg and Shamai \cite{Steinberg_Shamai_Marton}. With non-causal side information at the transmitter, Marton's innerbound becomes the convex hull of all rate pairs $(R_1, R_2)$ satisfying \cite{Steinberg_Shamai_Marton}: 
\begin{eqnarray*}
R_1&\leq& I(W,U; Y_1,S_{R1}) - I(W,U;S_T)\\
R_2&\leq& I(W,V; Y_2,S_{R2}) - I(W,V;S_T)\\
R_1+R_2&\leq&-[\max\{I(W;Y_1,S_{R1}),I(W;Y_2,S_{R2})\}-I(W;S_T)]_{+} \\
&&+ I(W,U;Y_1,S_{R1})-I(W,U;S_T)+I(W,V;Y_2,S_{R2})-I(W,V;S_T)-I(U;V|W,S_T)
\end{eqnarray*}
for all $P(U,V,W,X|S_T)$.

\subsection{Genie Bounds and the Relative Advantage of Non-causal Side Information over Causal Side Information}
Since the capacity region is not known for the general broadcast channel, there are two interesting alternatives to consider. As a first alternative, we could limit our attention to the special cases where the capacity region is known, such as the degraded broadcast channel. However, since our interest is to bound the \emph{sum} capacity improvement with causal and non-causal side information, the degraded case becomes trivial. For the degraded broadcast channel the sum capacity is the single user capacity of the stronger user, and therefore all the single user results obtained in previous sections apply. 

Another possibility is to limit our attention to Marton's innerbound region instead of the actual capacity region. In that case, the natural question is whether the extensions of Marton's innerbound for causal and non-causal side information become identical when the side information at the transmitter is also available at both the receivers. While this result was shown to be true for the general multiple access channel, it is interesting that the corresponding extension for the broadcast channel turns out to be less general.
\begin{observation} \label{obs1}
The extensions of Marton's innerbound with causal and non-causal side information (as described above) are identical if the following condition is satisfied:
\begin{equation*}
I(U;S_T|W)+I(V;S_T|W)=I(U,V;S_T|W)
\end{equation*}
\end{observation}
Intuitively, one can interpret the condition as follows. The common state information is provided by $W$ and therefore conditioned on $W$, the auxiliary random variables $U$ and $V$ should provide independent information about the state $S_T$. It is not known if this condition actually makes the achievable region smaller. A sketch of the proof of this observation is provided in the Appendix.

Following our previous results, the next question is to bound the maximum capacity benefit of genie bits. A straightforward extension of the single user result leads to the following bound: If a genie provides $G_1$ bits of side information to receiver 1 and $G_2$ bits of side information to receiver 2, the sum capacity can not improve by more than $G_1+G_2$ bits. The proof is straightforward since the improvement in the individual rates $R_1$, $R_2$ is bounded by $G_1$, $G_2$ respectively.
\subsection{Example: Broadcast Channel}
Consider a two user fading broadcast channel where the users' channels experience i.i.d. block fading (block size $N$) given by $h_1$ and $h_2$ respectively. 
\begin{eqnarray*}
Y_1 &=& h_1 X + n_1\\
Y_2 &=& h_2 X + n_2
\end{eqnarray*}
where $n_1, n_2 \sim {\mathcal{N}}(0,1)$ are additive white Gaussian noise terms and a transmit power constraint of $P$ is in place.
It is assumed that each user knows only his own channel, while the transmitter knows only the side information variable $S_T$ ($S_T=1$ if $h_1>h_2$ and $0$ otherwise). While in a degraded broadcast channel it is well known that the sum rate maximizing policy is to transmit only to the strongest user, the problem faced here is that a receiver does not know when it has the stronger channel compared to the other receiver's channel. Clearly, if a genie provides $S_T$ to both receivers then the sum capacity of this channel is $C_0=C_\Sigma(S_T,(h_1,S_T),(h_2,S_T))=\mbox{E}\log(1+P\max(|h_1|^2,|h_2|^2))$. Since the genie's bits can not increase capacity by more than $2$ bits (one bit to each receiver), the sum capacity (with causal or non-causal side information) can be bounded as:
\begin{eqnarray*}
C_0\geq  C_\Sigma^{\mbox{\tiny non-causal}}(S_T,h_1,h_2)
\geq C_\Sigma^{\mbox{\tiny causal}}(S_T,h_1,h_2)\geq C_0-2/N.
\end{eqnarray*}
\section{The Relay Channel}
The discrete memoryless relay channel with side information consists of source input alphabet $\mathcal{X}_S$, relay input alphabet $\mathcal{X}_R$, state alphabet $\mathcal{S}_S, \mathcal{S}_R, \mathcal{S}_D$ at the source (and relay and destination (respectively), relay output alphabet $\mathcal{Y}_R$, the destination output alphabet $\mathcal{Y}_D$, and a probability transition function $P(Y_R, Y_D|X_S, X_R, S_S, S_R, S_D)$.

Our first goal is to extend the result of Theorem \ref{theorem:cnc} to the relay channel. For the general relay channel, the capacity is unknown. For the special case of the degraded relay channel with no side information the capacity was found by \cite{Gamal_Relay}. In recent work, Sigurjonsson and Kim \cite{Sigurjonsson_Kim} combine Shannon's proof technique \cite{Shannon1958} with Cover and El Gamal's \cite{Gamal_Relay} relay coding theorems to determine the capacity of the degraded relay channel with causal side information for the special case $S_R=S_S=S$, i.e. the same state information is available to both the source and the relay. From the perspective of  Theorem \ref{theorem:cnc} we are interested in the scenario where the side information available to the source and relay is also available to the destination. Mathematically, we define the physically degraded channel with (causal or non-causal) side information where $S_R=S_S=S=f(S_D)$ as follows:
\begin{equation}\label{eq:degdef}
P(Y_D, Y_R | X_S, X_R, S, S_D)  = P(Y_R|X_S,X_R,S)P(Y_D|Y_R,X_R,S_D).
\end{equation}

\begin{figure}[h]
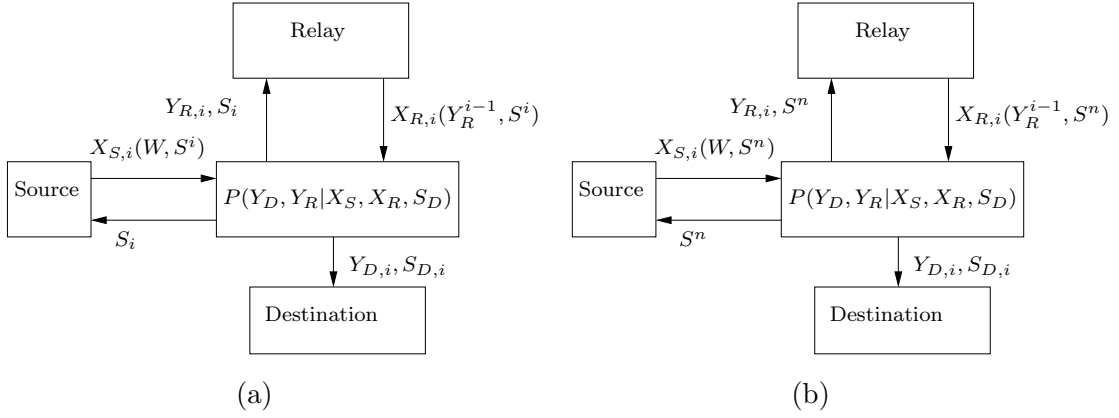

\[\begin{array}{lcl}
\input{relaycausal.pstex_t}&~~~~~~&\input{relaynoncausal.pstex_t}\\
\mbox{~~~~~~~~~~~~~~~~~~~~~~~~(a)} && \mbox{~~~~~~~~~~~~~~~~~~~~~~~(b)}\\
\end{array}
\]
\caption{Relay channel with (a) Causal and (b) Non-causal Side Information}\label{fig:relay}
\end{figure}

Figure \ref{fig:relay} illustrates the relay channel with side information.
The following theorem extends the result of Theorem \ref{theorem:cnc} to the relay channel.
\begin{theorem}\label{theorem:cncrelay}
The capacity of the physically degraded relay channel described above with causal or non-causal side information is given by:
\begin{eqnarray}
C^{\mbox{\tiny causal}}=C^{\mbox{\tiny non-causal}}
= \max_{P(X_S,X_R|S)}\min\left[I(X_S,X_R;Y_D|S_D),I(X_S;Y_R|X_R,S)\right]\nonumber
\end{eqnarray}
\end{theorem}
\proof: [Achievability] Achievability follows directly from the idea of multiplexed codebooks. Separate codebooks are designed for each possible state $S$. As the states are revealed in a causal fashion, the transmitter and the relay switch between the corresponding codebooks. Since the receiver also knows $S$ (a deterministic function of $S_D$), the receiver makes the corresponding switch as well. In this manner, for each state $S=s$, the capacity
\begin{eqnarray*}
C_s = \max_{P(X_S,X_R|S=s)}\min\left[I(X_S,X_R;Y_D|S_D, S=s),I(X_S;Y_R|X_R,S=s)\right]\nonumber
\end{eqnarray*}
is achieved. Averaging over $S$ we obtain the capacity expression of Theorem \ref{theorem:cncrelay}. Note that codebook multiplexing requires only causal side information. What is achievable with causal side information is also achievable with non-causal side information. \hfill\QED

\proof: [Converse] 
\begin{eqnarray*}
I(W;Y_D^n, S_D^n)&\leq& I(W;Y_D^n,Y_R^n|S_D^n)\\
&=& \sum_{i=1}^n H(Y_{D,i},Y_{R,i}|Y_D^{i-1},Y_R^{i-1},S_D^n)-H(Y_{D,i},Y_{R,i}|Y_D^{i-1},Y_R^{i-1},S_D^n, W)\\
&=& \sum_{i=1}^n H(Y_{D,i},Y_{R,i}|Y_D^{i-1},Y_R^{i-1},S_D^n, X_{R,i})-H(Y_{D,i},Y_{R,i}|Y_D^{i-1},Y_R^{i-1},S_D^n, W, X_{R,i}, X_{S,i})\\
&\leq& \sum_{i=1}^n H(Y_{D,i},Y_{R,i}| X_{R,i}, S_{D,i})-H(Y_{D,i},Y_{R,i}|S_{D,i} X_{R,i}, X_{S,i})\\
&=& \sum_{i=1}^n I(X_{S,i};Y_{D,i},Y_{R,i}|X_{R,i},S_{D,i})
\end{eqnarray*}
Using Fano's inequality and the time sharing variable to get a single letter characterization we get
\begin{eqnarray*}
R&\leq& I(X_S;Y_D,Y_R|X_R,S_D) +\epsilon_n\\
&=& I(X_S;Y_R|X_R,S_D)+ I(X_S;Y_D|X_R,S_D,Y_R)+\epsilon_n
\end{eqnarray*}
However, conditioned on $S_D$, the degraded channel condition implies that $X_S, (Y_R, X_R), Y_D$ is a Markov chain, i.e. $I(X_S;Y_D|X_R,S_D,Y_R)=0$ and we have,
\begin{eqnarray*}
R&\leq & I(X_S;Y_R|X_R,S_D)+ \epsilon_n\\
&= & H(Y_R|X_R,S_D,S)-H(Y_R|X_R,S_D,X_S,S)+ \epsilon_n\\
&\leq & H(Y_R|X_R,S)-H(Y_R|X_R,X_S,S)+ \epsilon_n\\
&=& I(X_S;Y_R|X_R,S) +\epsilon_n \label{eq:gg}
\end{eqnarray*}
Similarly, we have
\begin{eqnarray}
I(W;Y_D^n,S_D^n) &=& \sum_{i=1}^n H(Y_{D,i}|Y_D^{i-1},S_D^n)-H(Y_{D,i}|Y_D^{i-1},S_D^n,W)\\
&\leq& \sum_{i=1}^n H(Y_{D,i}|S_{D,i})-H(Y_{D,i}|Y_D^{i-1},S_D^n,W,Y_{R}^{i-1})\\
&=& \sum_{i=1}^n H(Y_{D,i}|S_{D,i})-H(Y_{D,i}|Y_D^{i-1},S_D^n,W,Y_{R}^{i-1},X_{S,i},X_{R,i})\\
&=& \sum_{i=1}^n H(Y_{D,i}|S_{D,i})-H(Y_{D,i}|S_{D,i},X_{S,i},X_{R,i})\\
&=& \sum_{i=1}^n I(X_{S,i},X_{R,i};Y_{D,i}|S_{D,i})\label{eq:g}
\end{eqnarray}
Combining (\ref{eq:g}) and (\ref{eq:gg}) with Fano's inequality we have,
\begin{eqnarray*}
R\leq \min\left[I(X_S,X_R;Y_D|S_D),I(X_S;Y_R|X_R,S)\right] +\tilde\epsilon_n 
\end{eqnarray*}
and the converse proof is complete.\hfill\QED
\subsection{Genie Bound and the Relative Advantage of Non-causal Side Information over Causal Side Information}
For the relay channel with only one destination, the result of Theorem \ref{theorem:genie} applies with the identical proof. In other words, one bit of genie information at the receiver does not improve capacity by more than one bit. Combining with the result of Theorem \ref{theorem:cncrelay} we have the following result:
For the physically degraded relay channel with side information defined by (\ref{eq:degdef}) (in general without the constraint $S=f(S_D)$) the difference between capacity with non-causal and causal side information at the transmitter is bounded as: $C^{\mbox{\tiny non-causal}}-C^{\mbox{\tiny causal}}\leq H(S|S_D)$.
\section{Conclusion}
Side information is a crucial factor in determining the capacity of a channel. Previous research has shown that for a single user memoryless channel even perfect feedback does not result in a capacity advantage. However, we find that even one bit of causal side information at the transmitter can increase capacity by an unbounded amount. The contrasting results are due to the fact that perfect feedback can only provide information about past channel states, whereas causal side information can  provide information about the current channel state. Thus, for the transmitter the difference between the past state information and present state information is very significant. We further explore the relative advantage of knowing the future channel states (non-causal side information) relative to the knowledge of only the present channel state (causal side information). For a single user, we find that the knowledge of future channel states can increase capacity only if the state information is not available to the receiver. In other words, non-causal side information has no capacity benefit over causal side information when the side information is available to the receiver as well. Note that for the receiver it does not matter if the side information is made available causally or non-causally, because no delay constraint is assumed in the decoding operation for capacity results. We evaluate the benefits of receiver side information in the form of a genie bound, which states quite simply that the capacity advantage from one genie bit of side information at the receiver can not exceed one bit. This gives us a bound on the capacity benefit of non-causal side information at the transmitter over causal side information in the form of the number of bits required to inform the receiver of the transmitter side information that is not already available to the receiver. 

We follow single user results with multiuser scenarios. All the results are found to extend to the general multiple access channel as well. For the broadcast channel, the extension is less general. In particular, for the broadcast channel it is not clear (even in terms of Marton's innerbound) if there is any benefit of non-causal side information over causal side information when both receivers know all the transmitter side information. This equivalence is established only subject to an interesting constraint. Finally, we show that the single user results extend to the degraded relay channel. Examples of random access channels are provided throughout as illustrations of the capacity bounds. While the MAC layer overheads in practical systems can be excessive the capacity results show that such overheads can be nearly eliminated through clever coding across multiple access blocks.

$~$\\
\appendix
\noindent{\Large\bf Appendix}
\section{Sketch of Proof of Observation \ref{obs1}}
When $S_T=f_1(S_{R1})=f_2(S_{R2})$, the following simplifications can be made, as in the proof of Theorem \ref{theorem:cnc}:
\begin{eqnarray*}
I(W,U;Y_1,S_{R1})-I(W,U;S_T) &=& I(W,U;Y_1|S_{R1}).\\
I(W,V;Y_2,S_{R2})-I(W,V;S_T) &=& I(W,V;Y_2|S_{R2}).\\
\max\{I(W;Y_1,S_{R1}),I(W;Y_2,S_{R2})\}-I(W,U;S_T) &=& \max\{I(W;Y_1|S_{R1}),I(W;Y_2|S_{R2})]\\
\end{eqnarray*}
and finally
\begin{eqnarray*}
&&-\max\{I(W;Y_1|S_{R1}),I(W;Y_2|S_{R2})]+ I(W,U;Y_1|S_{R1})+ I(W,V;Y_2|S_{R2}) \\
&=&\min\{I(W;Y_1| S_{R1})+I(W;Y_2|S_{R2})\}+I(U;Y_1|S_{R1},W)+I(V;Y_2|S_{R2},W)
\end{eqnarray*}

Using the simplifications it is easy to see that the causal case and the non-causal case are identical if $I(U;V|W,S)=I(U;V|W)$. But,
\begin{eqnarray*}
I(U;V|W,S)=I(U;V|W)+I(U,V;S|W)-I(U;S|W)-I(V;S|W)
\end{eqnarray*}
and thus we have the condition of observation \ref{obs1}.\hfill\QED
\bibliography{Thesis}

\begin{thebibliography}{10}

\bibitem{Shannon1958}
C.~E. Shannon, ``Channels with side information at the transmitter,'' in {\em
  IBMJ. Res. Develop}, pp.~289--293, 1958.

\bibitem{Kusnetsov_Tsybakov}
A.~Kusnetsov and B.~Tsybakov, ``Coding in a memory with defective cells,'' {\em
  Prob. Pered. Inform.}, vol.~10, pp.~52--60, April-June 1974.

\bibitem{Gelfand_Pinsker}
S.~I. Gelfand and M.~Pinsker, ``Coding for channel with random parameters,'' in
  {\em Problems of Control and Information Theory}, pp.~19--31, 1980.

\bibitem{Heegard_Gamal}
C.~Heegard and A.~{El Gamal}, ``On the capacity of computer memories with
  defects,'' {\em IEEE Trans. Inform. Theory}, vol.~29, pp.~731--739, Sep.
  1983.

\bibitem{Rosenzweig_Steinberg_Shamai}
A.~Rosenzweig, Y.~Steinberg, and S.~Shamai, ``On channels with partial channel
  state information at the transmitter,'' {\em IEEE Transactions on Information
  Theory}, vol.~51, pp.~1817--1830, May 2005.

\bibitem{Cemal_Steinberg}
Y.~Cemal and Y.~Steinberg, ``The multiple-access channel with partial state
  information at the encoders,'' {\em IEEE Trans. Inform. Theory}, vol.~51,
  pp.~3992--4003, Nov. 2005.

\bibitem{Cover_Chiang}
T.~M. Cover and M.~Chiang, ``Duality between channel capacity and rate
  distortion with two-sided state information,'' {\em IEEE Trans. Inform.
  Theory}, vol.~48, pp.~1629--1638, June 2002.

\bibitem{Jindal_Vishwanath_Goldsmith}
N.~Jindal, S.~Vishwanath, and A.~Goldsmith, ``On the duality of {Gaussian}
  multiple-access and broadcast channels,'' in {\em Proceedings of Int. Symp.
  Inform. Theory}, p.~500, June 2002.

\bibitem{Dimacs2}
N.~Jindal, S.~Vishwanath, S.~Jafar, and A.~Goldsmith, ``Duality, dirty paper
  coding and capacity for multiuser wireless channels,'' in {\em Proceedings of
  DIMACS workshop on Signal Processing for Wireless Transmissions}, October
  2002.

\bibitem{mimo_bc_journal}
S.~Vishwanath, N.~Jindal, and A.~Goldsmith, ``Duality, achievable rates, and
  sum-rate capacity of {MIMO} broadcast channels,'' {\em IEEE Trans. Inform.
  Theory}, pp.~2895--2909, Oct. 2003.

\bibitem{Steinberg_Merhav}
Y.~Steinberg and N.~Merhav, ``Identification in the presence of side
  information with application to watermarking,'' {\em IEEE Trans. Inform.
  Theory}, vol.~47, pp.~1410--1422, May 2001.

\bibitem{Costa}
M.~Costa, ``Writing on dirty paper,'' {\em IEEE Trans. Inform. Theory},
  vol.~29, pp.~439--441, May 1983.

\bibitem{Kim_Sutivong_Sigurjonsson}
Y.~{H. Kim}, A.~Sutivong, and S.~Sigurjonsson, ``Multiple user writing on dirty
  paper,'' {\em Proc. of IEEE Int. Symp. Information Theory}, p.~534, June/July
  2004.

\bibitem{Steinberg_Shamai_BC}
Y.~Steinberg and S.~Shamai, ``Achievable rates for the broadcast channel with
  states known at the transmitter,'' in {\em Proceedings of IEEE Int. Symp. on
  Info. Theory}, Sep. 2005.

\bibitem{Yu_Cioffi}
W.~Yu and J.~M. Cioffi, ``Sum capacity of a {Gaussian} vector broadcast
  channel,'' in {\em Proceedings of Int. Symp. Inform. Theory}, p.~498, June
  2002.

\bibitem{Viswanath_Tse}
P.~Viswanath and D.~N. Tse, ``Sum capacity of the multiple antenna {Gaussian}
  broadcast channel,'' in {\em Proceedings of Int. Symp. Inform. Theory},
  p.~497, June 2002.

\bibitem{Dimacs}
S.~Vishwanath, G.~Kramer, S.~Shamai, S.~Jafar, and A.~Goldsmith, ``Capacity
  bounds for {G}aussian vector broadcast channels,'' in {\em Proceedings of
  DIMACS workshop on Signal Processing for Wireless Transmissions}, October
  2002.

\bibitem{DimacsViswanath}
P.~Viswanath and D.~Tse, ``On the capacity of the multiple antenna broadcast
  channel,'' in {\em Proceedings of DIMACS workshop on Signal Processing for
  Wireless Transmissions}, October 2002.

\bibitem{Weingarten_Steinberg_Shamai}
H.~Weingarten, Y.~Steinberg, and S.~Shamai, ``The capacity region of the
  {G}aussian {MIMO} broadcast channel,'' in {\em Proceedings of Conference on
  Information Sciences and System s}, March 2004.

\bibitem{Salehi}
M.~Salehi, ``Capacity and coding for memories with real-time noisy defect
  information at encoder and decoder,'' {\em Proc. Inst. Elec. Eng. Pt. I},
  vol.~139, pp.~113--117, April 1992.

\bibitem{Caire_Shamai_CSI}
G.~Caire and S.~Shamai, ``On the capacity of some channels with channel state
  information,'' {\em IEEE Trans. Inform. Theory}, vol.~45, pp.~2007--2019,
  Sept 1999.

\bibitem{Das_Narayan}
A.~Das and P.~Narayan, ``Capacities of time-varying multiple-access channels
  with side information,'' {\em IEEE Trans. Inform. Theory}, vol.~48,
  pp.~4--25, Jan. 2002.

\bibitem{Sigurjonsson_Kim}
S.~Sigurjonsson and Y.~Kim, ``On multiple user channels with causal state
  information at the transmitters,'' {\em cs.IT/0508096}, 2005.
\newblock Download available at http://arxiv.org.

\bibitem{Cover_Thomas}
T.~M. Cover and J.~A. Thomas, {\em Elements of Information Theory}.
\newblock Wiley, 1991.

\bibitem{Chiang_Cover}
M.~Chiang and T.~Cover, ``Unified duality between channel capacity and rate
  distortion with state information,'' in {\em Proceedings of IEEE Int. Symp.
  Inform. Theory}, p.~301, June 2001.

\bibitem{Syed_cognitive}
S.~Jafar and S.~Srinivasa, ``Capacity limits of cognitive radio with
  distributed and dynamic spectral activity,'' {\em cs.IT/0509077}, 2005.
\newblock Download available at http://arxiv.org.

\bibitem{Marton}
K.~Marton, ``A coding theorem for the discrete memoryless broadcast channel,''
  {\em IEEE Trans. Inform. Theory}, vol.~25, pp.~306--311, May 1979.

\bibitem{Gamal_Meulen}
A.~{El Gamal} and E.~{van der Meulen}, ``A proof of {Marton's} coding theorem
  for the discrete memoryless broadcast channel,'' {\em IEEE Trans. Inform.
  Theory}, vol.~27, pp.~120--122, Jan. 1981.

\bibitem{Syed_CTW}
S.~Jafar, ``Degrees of freedom in distributed {MIMO} communication,'' {\em IEEE
  Communication Theory Workshop}, June 2005.

\bibitem{Steinberg_Shamai_Marton}
Y.~Steinberg and S.~Shamai, ``Achievable rates for the broadcast channel with
  state known at the transmitter,'' in {\em Proceedings of IEEE Int. Symp.
  Inform. Theory}, 2005.

\bibitem{Gamal_Relay}
T.~Cover and A.~E. Gamal, ``Capacity theorems for the relay channel,'' {\em
  IEEE Trans. Inform. Theory}, vol.~25, pp.~572--584, Sept. 1979.

\end{thebibliography}
\end{document}